\documentclass[fleqn,twoside]{article}
\usepackage{espcrc2}
\usepackage{graphicx}
\usepackage[figuresright]{rotating}

\newcommand{\AmS}{{\protect\the\textfont2
     A\kern-.1667em\lower.5ex\hbox{M}\kern-.125emS}}
\title{What is the order of the deconfining phase transition?}
\author{A. Di Giacomo\address[MCSD]{Dipartimento di Fisica,
Universit\`a di Pisa,\\
          I.N.F.N. Sezione di Pisa}%
          }
\begin{document}
\begin{abstract}
The techniques are discussed by which the order of the
deconfining phase transition is investigated on the lattice. QCD
with $N_f=2$ is a special case, which can provide information on the
mechanism of confinement.
\vspace{1pc}
\end{abstract}


\maketitle

\section{Introduction}
Quarks and gluons have never been observed as free particles, a
phenomenon known as color confinement.

A widely accepted idea in the 1960's was that a limiting temperature
exists for hadronic matter\cite{Hagedorn}, the Hagedorn temperature $T_H$:
increasing the energy density just below $T_H$ would result in production of
particles, rather than in an increase of temperature. In
ref.\cite{CP} it was suggested that a phase transition could exist
to a plasma of  quarks and gluons (Deconfining Transition).

Big experiments to explore this possibility in heavy ions collisions have
not yet given conclusive results\cite{QM02}.

The only existing source of information are Montecarlo simulations of the
theory on the Lattice.

The partition function of a system of fields at temperature T
\begin{equation}
                Z = Tr\{ \exp (- H/T)\}
\end{equation}
is equal to the Euclidean Feynmann integral of the theory extending in
time from 0 to $1/T$, with periodic boundary conditions in time  for bosons,
antiperiodic for fermions. In the Lattice formulation this corresponds
to simulate the theory on a lattice   $N_s^3\times N_t$, with the
spatial extension
$N_s \gg N_t$, the time extension.
     The temperature is   $T = 1/ a(\beta,m_q) N_t$, where    $\beta  =
2N/g^2$ and $a$ is the
lattice spacing in physical units.

     To explore the deconfining transition a signature for
confinement-deconfinement must be defined.
     In quenched QCD ($N_f=0$ , no dynamical quarks) this is done as follows.
Consider the Polyakov line
\begin{equation}
L(\vec x) = P\exp\left( i \int_0^{1/T} A_0(\vec x, t) dt\right) \label{eq2}
\end{equation}
The correlator
\begin{equation}
G(\vec r) = \langle L(\vec r)  L^\dagger(0)\rangle\label{eq3}
\end{equation}
is related to the static $q\bar q$ potential by the relation
\begin{equation}
    V(\vec r) = - T \ln G(\vec r)\label{eq4}
\end{equation}
On the other hand at large distances cluster property requires
\begin{equation}
    G(\vec r) \mathop\simeq_{r\to\infty}
    c\exp(-\frac{\sigma}{T} r) + \left|\langle L\rangle\right|^2
    \label{eq5}
\end{equation}
When $\langle L\rangle=0$   eq(\ref{eq5}) gives a linearily rising
potential at large distances, $V=\sigma r$ which is
taken as a definition of confinement. When  $\langle L\rangle\neq 0$,
$V(r) \to   const $ and there is no
confinement. $\langle L\rangle$    is an order parameter for
confinement,the corresponding
symmetry being $Z_3$.

A deconfining phase transition is indeed observed in lattice
simulations, corresponding to the above definition , both in $SU(2)$ and in
$SU(3)$ pure gauge theories:   $\langle L\rangle$   grows rapidly
from 0 to 1 by increasing $T$
at some $T_c$. In a finite lattice ,  however, the growth of
$\langle L\rangle$  is smooth:
no phase transition can take place in a finite system \cite{LY}. The
raise of   $\langle L\rangle$   becomes sharper and sharper as the
volume increases.

The infinite volume limit can be studied by a  technique used in
statistical mechanics \cite{F}, known as finite size scaling analysis.
The susceptibility of the order parameter  $\chi_L$ can be defined at a
given spatial size  $N_s$
\begin{equation}
    \chi_L(N_s) = \int d^3x\langle L(x) L^\dagger(0) -  L(0) L^\dagger(0)\rangle
    \label{eq6}
\end{equation}
$\chi_L$ gives a measure of the slope of the increase of $\langle
L(x)\rangle$   at $T_c$.
As $N_s\to\infty$
          it diverges with some critical index  $\gamma$ when approaching $T_c$
from below
    \begin{equation}
    \chi_L(N_s) \mathop\simeq_{T\to T_c^-} \tau^{-\gamma}\qquad \tau =
(1-\frac{T}{T_c})
    \label{eq7}
\end{equation}
The correlation length $\xi$ of the order parameter $\langle
L(x)\rangle$  also diverges
at the critical point with a critical index $\nu$
\begin{equation}
    \xi \mathop\simeq_{T\to T_c^-} \tau^{-\nu}
    \label{eq8}
\end{equation}
     Assuming that the ratio
    \[
    R(N_s) \equiv\frac{\chi_L(N_s)}{\chi_L(\infty)} =
f(\frac{a}{\xi},\frac{a N_s}{\xi})
    \]
    is an analytic function, as $T\to  T_c$,  $a/\xi\to 0$  and the
scaling law follows
\begin{equation}
R(N_s) \simeq f(0,\frac{a N_s}{\xi})
\label{eq9}
\end{equation}
or
\begin{equation}
\chi_L(N_s) \simeq N_s^{\gamma/\nu} \Phi_L(N_s^{1/\nu}\tau)
\label{eq10}
\end{equation}
The dependence on $N_s$ is dictated by the critical indices of the
transition. In particular the value of $\chi_L$ at the peak $\tau=0$ scales as
\begin{equation}
\chi_{L,peak}(N_s) \propto N_s^{\gamma/\nu}
\label{eq11}
\end{equation}
    A similar discussion gives for the specific heat $C_v$
\begin{equation}
C_v = C^0_v + N_s^{\alpha/\nu} \Phi_C(N_s^{1/\nu}\tau)
\label{eq12}
\end{equation}
with $\alpha$ the corresponding critical index and $C^0_v$ an additive term due
to the presence  of an additive renormalization.
     By studying numerically the dependence of  $\chi_L$ and $C_v$ on $N_s$ the
critical indices  $\alpha$, $\gamma$ ,$\nu$ can be determined and the order of
the transition and the universality class with them.
     A weak first order transition is a limiting case when $\alpha=\gamma=1$ and
$\nu=1/d$ with $d$ the number of spatial dimensions.

    For quenched $SU(2)$ it is found that the universality class is that of
the $3d$ ising model, and $\nu=.62$\cite{K}. For quenched $SU(3)$ the
transition is
weak first order \cite{TS} and $\nu=1/3$.

    An alternative order parameter  $\langle\mu\rangle$  is the vev of a
magnetically charged operator\cite{ref}. This operator will be described
in some detail in the next section for full QCD. A finite size scaling
analysis of
\[ \rho = \frac{d}{d\beta} \ln\langle\mu\rangle\]
gives results consistent with those obtained by using $\langle L\rangle$
and $\langle\mu\rangle\neq0$ for $T < T_c$, $\langle\mu\rangle=0$ for $T >
T_c$.

Since $\langle\mu\rangle \neq 0$ signals dual superconductivity, this
result can be
considered
as evidence that dual superconductivity of the vacuum is the mechanism of
confinement.
\section{Full QCD}
In the presence of dynamical quarks a clear way to define and to
detect confinement
does not exist. $\langle L\rangle$ is not a good order parameter, since the
corresponding symmetry $Z_3$ is broken by the quark coupling. In addition
string breaking is expected to occur: the static potential energy
     converts into dynamical $q-\bar q$ pairs at large distances so that the
potential is not linear any more  with the distance, even if there can be
confinement.

At $m_q=0$ chiral symmetry exists, which is spontaneously
broken at $T=0$, the pseudoscalar mesons being the corresponding Goldstone
particles and the quark condensate $\langle\bar \psi \psi\rangle$ the
order parameter. At some
critical temperature $T_c \sim 170$~Mev chiral symmetry is restored. It is
not well understood what is the interplay of chiral symmetry with
confinement. Moreover at $m_q \neq 0$ chiral symmetry is explicitely
broken and
   $\langle\bar \psi \psi\rangle$  is
not a good order parameter either.

However if one looks at
susceptibilities like  $\chi_L, \chi_{\bar\psi\psi}, C_v$     they
all show in numerical simulations a
maximum for a $T_c(m_q)$ as functions of $T$, along a line in the plane
$(m_q , T)$, which is assumed by convention as a critical line , with the
confined phase below
it and the deconfined phase above it\cite{KL}\cite{TS2}.

     We shall discuss a simplified model with $N_f=2$ and $m_u=m_d=m$, which is
semirealistic , but is of special interest as a probe of the theory.
The critical line is schetched in fig 1. for this model.

\begin{figure}
\includegraphics[scale=0.3]{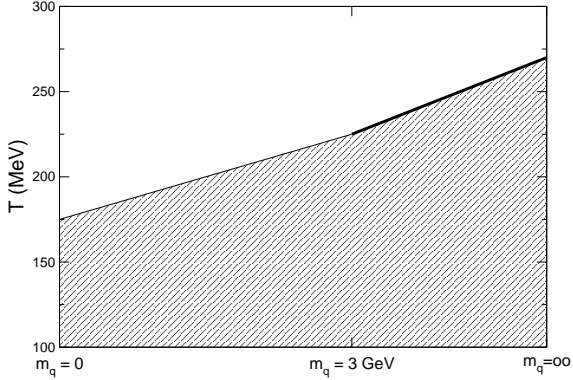}
\caption{Phase diagram of two flavor QCD}
\end{figure}

A renormalization group analysis of the chiral transition at $m=0$
\cite{PW} leads to the following predictions , based on the
hypothesis that the Golstone particles are the relevant degrees of
freedom at the transition. For $N_f=3$ or $2+1$ the transition is first order,
and such is expected to be for $m\neq 0$. For $N_f=2$ the transition is first
order both at $m =0$ and at $m\neq0$ if the axial $U(1)$ broken by
the anomaly is
restored at a lower  temperature than chiral symmetry,it is second order
if the anomaly persists up to the chiral transition, and in that case the
line in fig.1 is a crossover . There exists no precise numerical analysis
of the order of the transition , but for some reason the second
possibility is usually accepted.

      One can assume dual superconductivity of the vacuum as a criterion for
confinement. This criterion works in the quenched case, where it is
consistent with the one based on $\langle L\rangle$ \cite{ref}.
To be rigorous one should
prove that dual superconductivity implies confinement of any colored
particle . In the same way, however, one should prove that
$\langle L\rangle=0$ implies
confinement of any colored particle.

     Dual superconductivity is detected by a disorder parameter
$\langle \mu^a\rangle$, the
vev of an operator carrying non zero magnetic
charge\cite{ref}.A non zero vev of $\mu^a$ signals Higgs
breaking of
magnetic U(1) symmetry, or dual superconductivity. We shall  recall
below the definition and some properties of $\mu^a$.
     Numerical simulations show indeed that vacuum is a dual superconductor
in the region of fig~1 below the critical line , and is normal above
it\cite{ref} .

The operator $\mu^a$ is defined\cite{SUN} by the
orbits $U(x) \phi^a_{diag}U^\dagger(x)$ in the gauge group
of the symmetric space defined by the co-roots
\begin{equation}
    \phi_{diag}^a =
   diag\left(
\overbrace{\frac{N-a}{N},..,\frac{N-a}{N}}^{a},
\overbrace{-\frac{a}{N},..,-\frac{a}{N}}^{N-a}\right)
\label{eq13}\end{equation}
The definition is
\begin{equation}
\mu^a(\vec x,t) =
e^{
i\int d^3\vec y\,Tr\left(\phi^a(\vec y,t)\vec E(\vec y,t)\right)
\vec b_\perp(\vec x -\vec y)
}
\label{eq14}\end{equation}
Each choice of $U(x)$ defines an abelian projection\cite{tH81}, and
with it  monopoles.
The operator $\mu^a$ creates a monopole in the given abelian projection.
It can be shown\cite{D}\cite{DP}, however, that the property of
$\langle\mu^a\rangle$
of being non-zero or zero is abelian projection independent,if the
number density of monopoles is finite. Numerical simulations show
that this is indeed the case.
If one looks at the distribution of the difference of eigenvalues of
operators in the adjoint representation on the lattice sites, the
number of sites in which that difference vanishes is zero.
That difference is equal to zero at the locations of monopoles.
A large sample of configurations, lattice spacings, and operators
have been studied.

    A finite size scaling analysis allows to determine the
critical indices and the order of the transition\cite{referenza2}
     The problem has two scales, the correlation length and the quark mass
so that neglecting the ratio of the lattice spacing to the correlation
length,
\begin{equation}
    \langle\mu\rangle = f(\frac{N_s}{\xi}, m N_s^{y_h})\label{eq13}
    \end{equation}
where $y_h$ is the anomalous dimension of $m$ . By choosing values of $m$
and
$N_s$ such that the second argument is constant a scaling law follows for
the susceptibility
\[ \rho= \frac{d}{d\beta} \ln\langle \mu\rangle  \qquad
    \rho= N_s^{1/\nu} \Phi(N_s \tau^{1/\nu})\]
whence $\nu$ can be extracted. The result is consistent with a first order
thansition ($\nu=1/3$)\cite{PICA}.

This determination should be consistent with the
scaling of the specific heat eq(12). Preliminary data show that this is the
case indeed\cite{PICA}. Further numerical work is on the way to put
the result on a
firm basis. This would definitely legitimate the assumption of
$\langle\mu\rangle$ as an
order parameter for confinement, and confirm dual superconductivity as a
mechanism for color confinement. It would also imply\cite{PW} that either
axial
$U(1)$ symmetry is restored before chiral symmetry, or that Goldstone
particles are not the relevant degrees of freedom at the critical point.
The first possibility can also be tested on the lattice.

This work is partially supported by MIUR Progetto Teoria delle
interazioni fondamentali.
Thanks are due to my collaborators J.M. Carmona, L.Del Debbio, M.
D'Elia, B. Lucini, G. Paffuti, C. Pica for discussions .

\end{document}